\documentclass[
  prd,
  10pt,
  aps,
  onecolumn,
  tightenlines,
  superscriptaddress,
  amsmath,
  amssymb,
  amsfonts,
  nofootinbib,
  floatfix
]{revtex4-2}

\usepackage[utf8]{inputenc}
\usepackage[T1]{fontenc}
\usepackage[english]{babel}
\usepackage{lmodern}
\usepackage{microtype}
\usepackage{graphicx}
\usepackage{amsmath,amssymb,amsfonts}
\usepackage{bm}
\usepackage{braket}
\usepackage{float}
\usepackage{xcolor}

\graphicspath{{./}}
\newcommand{\ii}{\mathrm{i}}
\newcommand{\dd}{\mathrm{d}}

\begin{document}
\emergencystretch=1.5em

\title{Spin-dependent neutrino oscillations in torsion backgrounds:\\
A quantum-field-theoretic analysis}

\author{R. Serao}
\email{rserao@unisa.it}
\affiliation{Dipartimento di Fisica ``E.R. Caianiello'', Universit\`a di Salerno,
and INFN -- Gruppo Collegato di Salerno, Via Giovanni Paolo II 132,
84084 Fisciano (SA), Italy}

\author{G. De Maria}
\email{giuseppedemaria97@gmail.com}
\affiliation{DIME -- Sezione Metodi e Modelli Matematici, Universit\`a di Genova,
Via All'Opera Pia 15, 16145 Genova, Italy}

\author{S. Monda}
\email{s.monda5@studenti.unisa.it}
\affiliation{Dipartimento di Fisica ``E.R. Caianiello'', Universit\`a di Salerno,
and INFN -- Gruppo Collegato di Salerno, Via Giovanni Paolo II 132,
84084 Fisciano (SA), Italy}

\author{A. Quaranta}
\email{aniello.quaranta@unicam.it}
\affiliation{School of Science and Technology, University of Camerino, Via Madonna delle Carceri, Camerino, 62032, Italy}

\author{A. Capolupo}
\email{capolupo@sa.infn.it}
\affiliation{Dipartimento di Fisica ``E.R. Caianiello'', Universit\`a di Salerno,
and INFN -- Gruppo Collegato di Salerno, Via Giovanni Paolo II 132,
84084 Fisciano (SA), Italy}

\begin{abstract}
We study neutrino mixing in a background with spacetime torsion within the
quantum-field-theoretic formulation of flavor oscillations. Working in the
Einstein--Cartan framework and neglecting curvature, we quantize Dirac fields
in constant and linearly time-dependent axial-torsion backgrounds. A constant
spatial torsion component lifts the degeneracy between the two spin
orientations through spin-dependent effective masses and energies. In quantum
field theory this splitting modifies not only the oscillation frequencies but
also the amplitudes, because the Bogoliubov coefficients entering the flavor
operators depend on spin. The effect is largest at low momentum when the
torsion scale is comparable to the neutrino masses, while a dominant torsion
term suppresses the relative mass splittings and can inhibit flavor
conversion. We also discuss the induced spin dependence of the Dirac
$CP$ asymmetry and of the condensate densities in the flavor vacuum. The
results identify nonrelativistic neutrinos as the natural regime in which the
difference between the field-theoretic and quantum-mechanical descriptions is
most pronounced.
\end{abstract}

\maketitle

\section{Introduction}
\label{sec:introduction}

General relativity describes gravitation through the geometry of spacetime,
with the gravitational interaction encoded in the metric tensor and its
torsion-free Levi--Civita connection. Despite its remarkable theoretical and
observational success, general relativity is not the only geometrically
consistent description of gravity. A broad class of extended theories has been
developed in which additional gravitational degrees of freedom arise from
scalar fields, higher-order curvature invariants, or a more general affine
structure
\cite{Capozziello2011,BransDicke,Quiros2019,Kobayashi2019}.
Metric-affine formulations are especially relevant in this respect because the
metric and the affine connection are treated, in general, as independent
variables. Curvature, torsion, and nonmetricity can then be regarded as
distinct geometric structures entering the gravitational dynamics
\cite{CapozzielloVignolo2010,
CapozzielloHarkoKoivistoLoboOlmo2015}.

This broader geometric perspective also underlies the so-called geometric
trinity of gravity. General relativity describes gravitation in terms of
curvature, the teleparallel equivalent of general relativity employs torsion,
and the symmetric teleparallel formulation is based on nonmetricity. At the
level of the Einstein dynamics, these formulations may be related through
appropriate boundary terms, although their geometric foundations and natural
extensions need not be equivalent
\cite{Cai2016,CapozzielloDeFalcoFerrara2022,
CapozzielloCesareFerrara2025,ManciniTinoCapozziello2025}. It is important, however, not to identify the
torsion used in teleparallel gravity with the torsion appearing in the
Einstein--Cartan theory. In the former, torsion represents the gravitational
field in a curvature-free formulation, whereas in the latter it is the
antisymmetric part of a metric-compatible affine connection and is naturally
sourced by the intrinsic spin of matter.

The Einstein--Cartan theory provides the simplest extension of general
relativity in which this spin--torsion coupling is retained. It is obtained by
relaxing the assumption that the affine connection is symmetric and identifying
its antisymmetric component with spacetime torsion
\cite{Hehl1976,Shapiro2002}. Whereas curvature is associated with the
energy--momentum distribution, torsion is naturally related to spin density.
This distinction becomes particularly relevant for fermionic fields. Under
minimal coupling, only the totally antisymmetric, or axial, component of
torsion contributes directly to the Dirac equation. The resulting axial-vector
interaction can induce spin-dependent corrections to fermionic dispersion
relations, energy-level splittings, and spin precession
\cite{Cabral2021,Cirilo2019}. Torsion thus provides a
natural setting in which the microscopic spin structure of matter can affect
particle propagation.

Neutrinos are especially sensitive probes of weak modifications of their
propagation. Their small masses, macroscopic propagation distances, and
flavor-changing dynamics imply that even small corrections to their
single-particle energies can accumulate into appreciable changes of the
relative oscillation phases. In flat spacetime and in the relativistic regime,
neutrino oscillations are successfully described by the Pontecorvo framework,
in which flavor states are coherent superpositions of states with definite
masses
\cite{Bilenky1978,Bilenky1987,Fukuda1998}. Gravitational or matter
backgrounds may nevertheless alter the evolution of the mass eigenstates and
thereby modify the flavor-transition pattern. This possibility has motivated
the investigation of neutrino propagation in cosmological and astrophysical
geometries, where spacetime expansion, gravitational redshift, horizons, and
the local geometry of the propagation region may affect the oscillation
phases.

A complete analysis of these effects requires quantum field theory in curved
spacetime. In a generic curved geometry, the decomposition of a quantum field
into positive- and negative-frequency modes is not globally unique. Field
quantization therefore introduces a Bogoliubov structure that combines with
the Bogoliubov transformation intrinsically associated with particle mixing.
General quantum-field-theoretic oscillation formulae have been constructed for
fermionic fields on globally hyperbolic spacetimes and explicitly applied to
Friedmann--Lemaître--Robertson--Walker and Schwarzschild backgrounds
\cite{CapolupoLambiaseQuaranta2020,
CapolupoQuarantaSerao2023}. These analyses show that a gravitational
background may modify not only the relative phases, as in the usual
quantum-mechanical description, but also the oscillation amplitudes through
the curved-spacetime spinor overlaps and the corresponding Bogoliubov
coefficients. The flat-spacetime and ultrarelativistic expressions are
recovered in the appropriate limits.

Even in Minkowski spacetime, the quantum-field-theoretic formulation of
particle mixing is richer than the Pontecorvo construction. The transformation
between fields with definite masses and fields with definite flavors is
implemented at the operator level by a time-dependent mixing generator and
contains both particle--particle and particle--antiparticle contributions.
The massive and flavor representations become unitarily inequivalent in the
infinite-volume limit, and the physical flavor vacuum acquires a nontrivial
condensate structure
\cite{Blasone1995,Blasone2002,Blasone2001}. The exact
oscillation formulae consequently contain the usual low-frequency terms,
determined by differences of one-particle energies, together with genuine
field-theoretic contributions oscillating with sums of energies. These
corrections are strongly suppressed for ultrarelativistic particles but may
become relevant in the nonrelativistic regime, where the momentum is
comparable with the particle masses.

The role of coherence, environmental interactions, and nonclassical
correlations has also established a connection between neutrino physics and
quantum-information theory. Open-system methods and entanglement measures may
provide complementary tools for studying unconventional oscillation patterns,
decoherence, the possible Dirac or Majorana nature of neutrinos, violations of
fundamental symmetries, and interactions mediated by weakly coupled fields
\cite{SeraoTorreCapolupo2025QuantumInformation}. From this perspective, an
external torsion background represents another physical setting in which the
coherent evolution of a mixed quantum field is altered by additional
degrees of freedom. The present analysis is formulated in terms of exact
flavor charges and oscillation functions rather than explicit
information-theoretic observables, but the two descriptions address
complementary aspects of the same underlying quantum dynamics.

The condensate structure of the flavor vacuum also establishes a connection
between particle mixing, gravitation, and the dark sector. The expectation
value of the energy--momentum tensor on the flavor vacuum may behave as an
effective cosmological fluid. For mixed fermions in expanding backgrounds, it
has been shown to possess an equation of state compatible with dust and cold
dark matter
\cite{Capolupo2016,CapolupoCarloniQuaranta2022,
CapolupoMondaPisacaneQuarantaSerao2025}. The same framework has
been extended to static and spherically symmetric geometries, where the
semiclassical flavor-vacuum contribution generates, in the weak-field limit,
a correction to the Newtonian potential with possible consequences for
galactic dynamics
\cite{CapolupoCapozzielloPisacaneQuaranta2025}.

The low-momentum regime is particularly relevant because it is precisely there
that the quantum-field-theoretic corrections to the Pontecorvo description
are largest. Neutrino capture on tritium has consequently been proposed as a
possible probe of the cosmic neutrino background and of the condensate
structure associated with the flavor vacuum
\cite{CapolupoQuaranta2023}. Independent but complementary
strategies employ quantum-optical systems to search for weakly coupled dark
sectors. In particular, a single-arm interferometric configuration involving
spatially separated squeezing and antisqueezing operations has been proposed
to probe the interaction between photons and an ultralight scalar dark-matter
field
\cite{CapolupoPisacaneQuarantaSerao2026Interferometry}. These approaches
illustrate how phase-sensitive low-energy experiments may access interactions
that are otherwise extremely difficult to observe.

More generally, precision observables can also constrain light particles
beyond the Standard Model. One example is the proposed $X_{17}$ vector boson,
whose possible contributions to anomalous magnetic moments, the muonic Lamb
shift, and electroweak observables have been investigated in
Ref.~\cite{CapolupoQuarantaSerao2025X17}. Such particle-mediated interactions
are physically distinct from a geometrical torsion background. Their relevance
here is therefore methodological rather than dynamical: both cases illustrate
how small corrections to fermionic energies, phases, or precision observables
may reveal weakly coupled sectors that are absent from the minimal Standard
Model.

The inclusion of spacetime torsion provides a further and conceptually
distinct generalization of neutrino oscillation physics. Neutrino propagation
in torsional geometries has previously been considered mainly within
quantum-mechanical treatments
\cite{Adak2001,Fabbri2016}. In such descriptions, torsion modifies the
effective dispersion relations and hence the phases accumulated by the
different mass eigenstates. A quantum-field-theoretic treatment predicts an
additional effect. Since the torsion background changes the fermionic mode
functions and their spin dependence, it also changes the spinor overlaps
entering the Bogoliubov transformation between massive and flavor operators.
Torsion can therefore affect both the frequencies and the amplitudes of the
exact oscillation functions. The two spin sectors may obey different
oscillation formulae even when transitions involving an explicit spin flip
are neglected.

The full quantum-field-theoretic construction for constant and linearly
time-dependent torsion backgrounds was developed in
Ref.~\cite{CapolupoDeMariaMondaQuarantaSerao2024}, with subsequent concise
accounts and phenomenological discussions presented in
Refs.~\cite{
CapolupoMondaPisacaneQuarantaSerao2026Torsion}. The purpose of the present
contribution is to provide a compact and self-contained account of this
framework, emphasizing the distinction between phase corrections and the
genuinely field-theoretic deformation of the oscillation amplitudes. We first
review the geometrical origin of the Dirac--torsion coupling and the associated
spin-dependent mode solutions. We then construct the flavor fields and their
Bogoliubov coefficients, derive a representative exact three-flavor
oscillation formula, and discuss the implications for flavor conversion,
$CP$ violation, and the condensation densities of the flavor vacuum.
Particular attention is devoted to nonrelativistic neutrinos, for which the
departure from the quantum-mechanical description is expected to be most
pronounced.

\section{Spacetime torsion and Dirac fields}

\subsection{Riemann--Cartan geometry and the axial coupling}

A metric-compatible connection with torsion can be written as
\begin{equation}
\widetilde{\Gamma}^{\rho}_{\mu\nu}
=\Gamma^{\rho}_{\mu\nu}+K^{\rho}_{\mu\nu},
\qquad
T^{\rho}{}_{\mu\nu}
=\widetilde{\Gamma}^{\rho}_{\mu\nu}
-\widetilde{\Gamma}^{\rho}_{\nu\mu},
\label{eq:connection}
\end{equation}
where $\Gamma^{\rho}_{\mu\nu}$ is the Levi--Civita connection and
$K^{\rho}_{\mu\nu}$ is the contorsion tensor,
\begin{equation}
K_{\rho\mu\nu}
=\frac{1}{2}
\left(T_{\rho\mu\nu}+T_{\mu\nu\rho}-T_{\nu\rho\mu}\right).
\end{equation}
In four dimensions the torsion tensor admits the irreducible decomposition
\begin{equation}
T_{\rho\mu\nu}
=\frac{1}{3}\left(V_{\mu}g_{\rho\nu}-V_{\nu}g_{\rho\mu}\right)
-\frac{1}{6}\epsilon_{\rho\mu\nu\sigma}T^{\sigma}
+q_{\rho\mu\nu},
\label{eq:torsion-decomposition}
\end{equation}
where $V_{\mu}=T^{\rho}{}_{\mu\rho}$ is the vector trace,
$T^{\mu}=\epsilon^{\alpha\beta\gamma\mu}T_{\alpha\beta\gamma}$ is the
axial component, and $q_{\rho\mu\nu}$ is the remaining traceless tensor part
\cite{Shapiro2002}. For a minimally coupled Dirac field, only the axial
component enters the interaction directly.

The spinor covariant derivative is
\begin{equation}
\widetilde D_{\mu}\psi
=D_{\mu}\psi
+\frac{1}{4}K_{AB\mu}[\gamma^{A},\gamma^{B}]\psi,
\end{equation}
and the Dirac action becomes
\begin{align}
\widetilde S_D
={}&\int \dd^4x\sqrt{-g}
\left[
\frac{\ii}{2}
\left(\bar\psi\gamma^{\mu}D_{\mu}\psi
-D_{\mu}\bar\psi\gamma^{\mu}\psi\right)
-m\bar\psi\psi
\right]
\nonumber\\
&+3\int \dd^4x\sqrt{-g}\,T_{\mu}S^{\mu},
\qquad
S^{\mu}=\frac{1}{2}\bar\psi\gamma^{\mu}\gamma^{5}\psi.
\label{eq:dirac-action}
\end{align}
We treat torsion as an externally generated background and set the metric
curvature to zero in order to isolate its effect. The field equation then
reduces to
\begin{equation}
\left(\ii\gamma^{\mu}\partial_{\mu}-m\right)\psi
=-\frac{3}{2}T_{\rho}\gamma^{\rho}\gamma^{5}\psi.
\label{eq:dirac-torsion}
\end{equation}
This assumption is deliberately restricted: the analysis determines how a
given torsion profile affects neutrino mixing, but it does not specify the
astrophysical source or solve the coupled Einstein--Cartan field equations.

\subsection{Constant axial torsion}

Consider a constant axial torsion directed along the third spatial axis. The
Dirac equation reads
\begin{equation}
\ii\gamma^{\mu}\partial_{\mu}\psi
=m\psi-\frac{3}{2}T_{3}\gamma^{3}\gamma^{5}\psi.
\label{eq:constant-dirac}
\end{equation}
For momentum parallel to the torsion axis, the positive- and negative-energy
solutions have the same algebraic form as the free Dirac spinors, with the
replacement
\begin{equation}
\widetilde m_i^{\,r}=m_i+r\frac{3}{2}T^{3},
\qquad
E_{i,\bm k}^{r}
=\sqrt{\bm k^2+\left(\widetilde m_i^{\,r}\right)^2},
\qquad r=\pm.
\label{eq:effective-mass}
\end{equation}
The normalization factor is
$N_i^{r}=\sqrt{(E_{i,\bm k}^{r}+\widetilde m_i^{\,r})/(2E_{i,\bm k}^{r})}$.
The field expansion is therefore
\begin{equation}
\nu_i(x)
=\sum_r\int\frac{\dd^3k}{(2\pi)^{3/2}}
\left[
 u_{i,\bm k}^{r}(t)\alpha_{i,\bm k}^{r}
+v_{i,-\bm k}^{r}(t)\beta_{i,-\bm k}^{r\dagger}
\right]e^{\ii\bm k\cdot\bm x}.
\label{eq:field-expansion}
\end{equation}
Equation~\eqref{eq:effective-mass} makes the first physical consequence
transparent: torsion lifts the degeneracy between the two spin orientations.
Because the spinors depend on the shifted masses, their overlaps also acquire
a spin dependence.

\subsection{Linearly time-dependent torsion}

A nonstationary background can be treated analytically when the temporal
component $\breve T^{0}$ is constant and the spatial components depend only on
time. Writing $\breve T^{i}(t)=\alpha^{i}t$, and expanding in helicity
eigenspinors $\xi_{\lambda}(\hat{\bm p})$, the mode functions satisfy
\begin{equation}
\ii\partial_t
\begin{pmatrix}f_{\bm p,\lambda}\\g_{\bm p,\lambda}\end{pmatrix}
=
\begin{pmatrix}
 m-\eta\lambda\breve T^{i}(t)\hat p_i
 & p+\eta\lambda\breve T^{0}\\
 p+\eta\lambda\breve T^{0}
 &-m-\eta\lambda\breve T^{i}(t)\hat p_i
\end{pmatrix}
\begin{pmatrix}f_{\bm p,\lambda}\\g_{\bm p,\lambda}\end{pmatrix}.
\label{eq:time-matrix}
\end{equation}
Since the matrices at different times commute for constant $\breve T^0$, the
solution is obtained by ordinary exponentiation. Its two characteristic
quantities are
\begin{equation}
\omega_{i,p,\lambda}
=\sqrt{m_i^2+\left(p+\eta\lambda\breve T^0\right)^2},
\qquad
\exp\left[-\ii\eta\lambda\hat p_i
\int_0^t\dd\tau\,\breve T^i(\tau)\right]
=
\exp\left[-\frac{\ii}{2}\eta\lambda\alpha_i\hat p^i t^2\right].
\label{eq:time-solution}
\end{equation}
The first shifts the helicity-dependent dispersion relation; the second adds a
nonlinear dynamical phase. In the following we neglect transitions that flip
the spin. This restriction must be kept in mind when interpreting the
nonstationary results.

\section{Quantum field theory of neutrino mixing in a torsion background}

Let $\Psi_m^T=(\nu_1,\nu_2,\nu_3)$ denote the fields with definite masses and
$\Psi_f^T=(\nu_e,\nu_\mu,\nu_\tau)$ the flavor triplet. The usual PMNS rotation
can be implemented at the operator level by a time-dependent mixing generator,
\begin{equation}
\nu_\sigma^{\alpha}(x)
=I_\theta^{-1}(t)\nu_i^{\alpha}(x)I_\theta(t),
\qquad
I_\theta(t)=I_{23}(t)I_{13}(t)I_{12}(t),
\label{eq:generator}
\end{equation}
where $(\sigma,i)=(e,1),(\mu,2),(\tau,3)$. At finite volume, the flavor vacuum
is
\begin{equation}
\ket{0(t)}_f=I_\theta^{-1}(t)\ket{0}_m.
\label{eq:flavor-vacuum}
\end{equation}
The flavor annihilation operators are not simple rotations of the massive
operators. They also contain antiparticle creation operators, with
coefficients fixed by the overlaps of spinors with different masses. This is
the Bogoliubov structure responsible for the exact field-theoretic corrections
to the Pontecorvo formulae.

For constant torsion and $\bm k=(0,0,k)$, the nonvanishing coefficients are
diagonal in spin. Their moduli may be written as
\begin{align}
\left|\Xi_{ij;\bm k}^{rr}\right|
={}&N_i^rN_j^r
\left[
1+\frac{k^2}
{\left(E_{i,\bm k}^r+\widetilde m_i^{\,r}\right)
 \left(E_{j,\bm k}^r+\widetilde m_j^{\,r}\right)}
\right],
\label{eq:Xi}\\
\left|\chi_{ij;\bm k}^{rr}\right|
={}&N_i^rN_j^r k
\left[
\frac{1}{E_{j,\bm k}^r+\widetilde m_j^{\,r}}
-\frac{1}{E_{i,\bm k}^r+\widetilde m_i^{\,r}}
\right],
\label{eq:chi}
\end{align}
with phases
\begin{equation}
\Xi_{ij;\bm k}^{rr}(t)
=\left|\Xi_{ij;\bm k}^{rr}\right|
 e^{\ii(E_{j,\bm k}^r-E_{i,\bm k}^r)t},
\qquad
\chi_{ij;\bm k}^{rr}(t)
=\left|\chi_{ij;\bm k}^{rr}\right|
 e^{\ii(E_{j,\bm k}^r+E_{i,\bm k}^r)t}.
\label{eq:coeff-phases}
\end{equation}
Canonicity requires
$|\Xi_{ij;\bm k}^{rr}|^2+|\chi_{ij;\bm k}^{rr}|^2=1$.
For the time-dependent background, the corresponding coefficients are denoted
by $\Pi_{ij;\bm p}^{rr}(t)$ and $\Upsilon_{ij;\bm p}^{rr}(t)$ and are defined by
the same spinor inner products. They satisfy
\begin{equation}
\left|\Pi_{ij;\bm p}^{rr}(t)\right|^2
+\left|\Upsilon_{ij;\bm p}^{rr}(t)\right|^2=1.
\end{equation}
It is convenient to use the unified notation
\begin{equation}
(\Gamma,\Sigma)=
\begin{cases}
(\Xi,\chi),&\text{constant torsion},\\
(\Pi,\Upsilon),&\text{time-dependent torsion}.
\end{cases}
\label{eq:unified}
\end{equation}
In the ultrarelativistic limit, $\Gamma\to1$ and $\Sigma\to0$, and the usual
quantum-mechanical structure is recovered.

\section{Spin-dependent neutrino oscillations}

The flavor charges are obtained by normal ordering with respect to the flavor
vacuum,
\begin{equation}
::Q_{\nu_\sigma}(t)::
=\sum_r\int\dd^3k
\left[
\alpha_{\bm k,\nu_\sigma}^{r\dagger}(t)
\alpha_{\bm k,\nu_\sigma}^{r}(t)
-
\beta_{-\bm k,\nu_\sigma}^{r\dagger}(t)
\beta_{-\bm k,\nu_\sigma}^{r}(t)
\right].
\label{eq:flavor-charge}
\end{equation}
For a neutrino created with flavor $\rho$, momentum $\bm k$, and spin $r$, the
transition function is
\begin{equation}
\mathcal Q_{\nu_\rho\to\nu_\sigma}^{r,\bm k}(t)
={}_f\!\bra{\nu_\rho^{r,\bm k}}
::Q_{\nu_\sigma}(t)::
\ket{\nu_\rho^{r,\bm k}}_f.
\end{equation}
A representative exact result is the electron-neutrino survival function,
\begin{align}
\mathcal Q_{\nu_e\to\nu_e}^{r,\bm k}(t)
={}&1
-\sin^2(2\theta_{12})\cos^4\theta_{13}
\left[
|\Gamma_{12;\bm k}^{rr}|^2\sin^2(\Delta_{12;\bm k}^r t)
+|\Sigma_{12;\bm k}^{rr}|^2\sin^2(\Omega_{12;\bm k}^r t)
\right]
\nonumber\\
&-\sin^2(2\theta_{13})\cos^2\theta_{12}
\left[
|\Gamma_{13;\bm k}^{rr}|^2\sin^2(\Delta_{13;\bm k}^r t)
+|\Sigma_{13;\bm k}^{rr}|^2\sin^2(\Omega_{13;\bm k}^r t)
\right]
\nonumber\\
&-\sin^2(2\theta_{13})\sin^2\theta_{12}
\left[
|\Gamma_{23;\bm k}^{rr}|^2\sin^2(\Delta_{23;\bm k}^r t)
+|\Sigma_{23;\bm k}^{rr}|^2\sin^2(\Omega_{23;\bm k}^r t)
\right],
\label{eq:survival}
\end{align}
where
\begin{equation}
\Delta_{ij;\bm k}^r
=\frac{E_{j,\bm k}^r-E_{i,\bm k}^r}{2},
\qquad
\Omega_{ij;\bm k}^r
=\frac{E_{j,\bm k}^r+E_{i,\bm k}^r}{2}.
\label{eq:frequencies}
\end{equation}
The first term in each bracket is the deformation of the usual oscillation
contribution. The second is intrinsically field-theoretic and oscillates at the
sum of the one-particle energies. Torsion enters both pieces: it changes the
frequencies through Eq.~\eqref{eq:effective-mass} and the amplitudes through the
spin-dependent coefficients in Eqs.~\eqref{eq:Xi} and \eqref{eq:chi}.

This distinction is the main result. In a quantum-mechanical treatment one
sets $\Gamma=1$ and $\Sigma=0$; the spin orientation then changes only the
dispersion relation and hence the phase. In quantum field theory,
\begin{equation}
\mathcal Q_{\nu_\rho\to\nu_\sigma}^{+,\bm k}(t)
\neq
\mathcal Q_{\nu_\rho\to\nu_\sigma}^{-,\bm k}(t)
\end{equation}
through both the phase and the oscillation amplitude. Probability conservation
is preserved,
$\sum_{\sigma=e,\mu,\tau}\mathcal Q_{\nu_\rho\to\nu_\sigma}^{r,\bm k}=1$.

\section{Phenomenological implications}

\subsection{Constant torsion}

To illustrate the effect, we use the representative masses
\begin{equation}
m_1=10^{-3}\,\mathrm{eV},\qquad
m_2=9\times10^{-3}\,\mathrm{eV},\qquad
m_3=2\times10^{-2}\,\mathrm{eV},
\end{equation}
which reproduce mass-squared splittings of the observed order, together with
$\sin^2(2\theta_{12})=0.861$,
$\sin^2(2\theta_{23})=0.97$,
$\sin^2(2\theta_{13})=0.10$, and $\delta=\pi/4$. We take
$k=2\times10^{-2}\,\mathrm{eV}$ and
$|T^3|=2\times10^{-4}\,\mathrm{eV}$. These values are not a fit to a specific
astrophysical source; they are a benchmark chosen to expose the low-momentum
regime in which the field-theoretic corrections are visible.

\begin{figure}[htbp]
\centering
\includegraphics[width=0.47\textwidth]{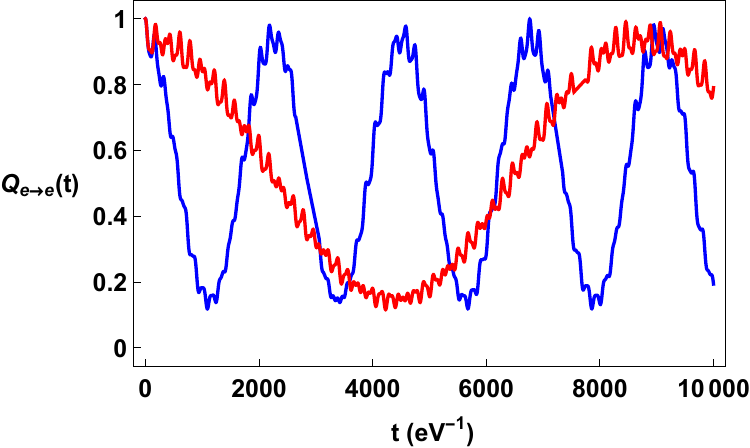}\hfill
\includegraphics[width=0.47\textwidth]{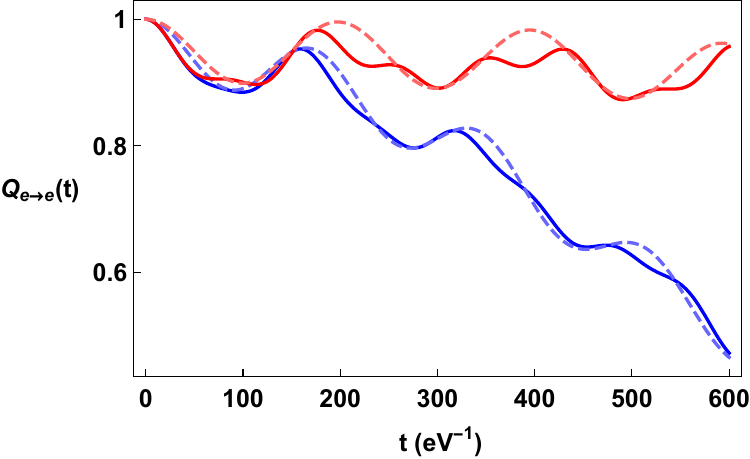}
\caption{Electron-neutrino survival in a constant torsion background. Left:
$\mathcal Q_{\nu_e\to\nu_e}^{+,\bm k}$ (blue) and
$\mathcal Q_{\nu_e\to\nu_e}^{-,\bm k}$ (red). Right: short-time detail and
comparison with the corresponding quantum-mechanical curves (dashed). The
benchmark uses $T^3=2\times10^{-4}\,\mathrm{eV}$ and
$k=2\times10^{-2}\,\mathrm{eV}$.}
\label{fig:constant-survival}
\end{figure}

\begin{figure}[htbp]
\centering
\includegraphics[width=0.47\textwidth]{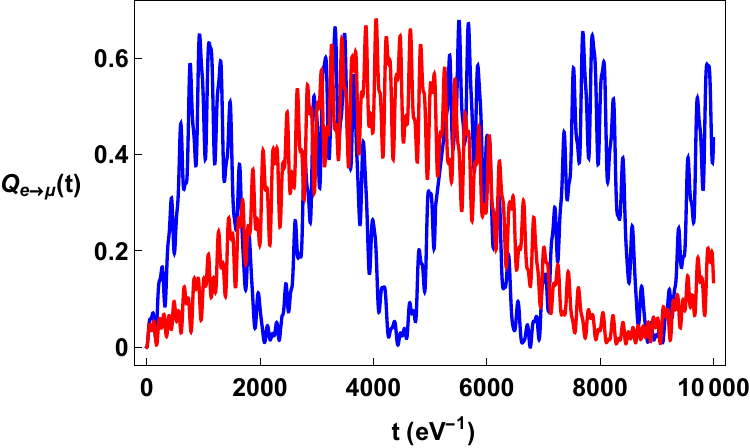}\hfill
\includegraphics[width=0.47\textwidth]{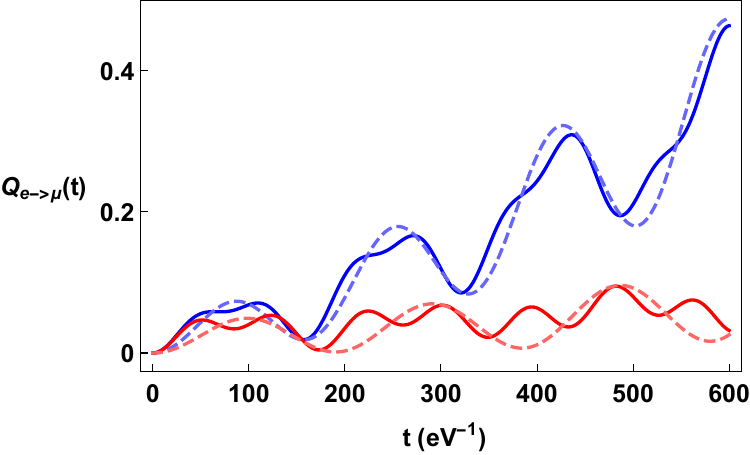}\\[0.8em]
\includegraphics[width=0.47\textwidth]{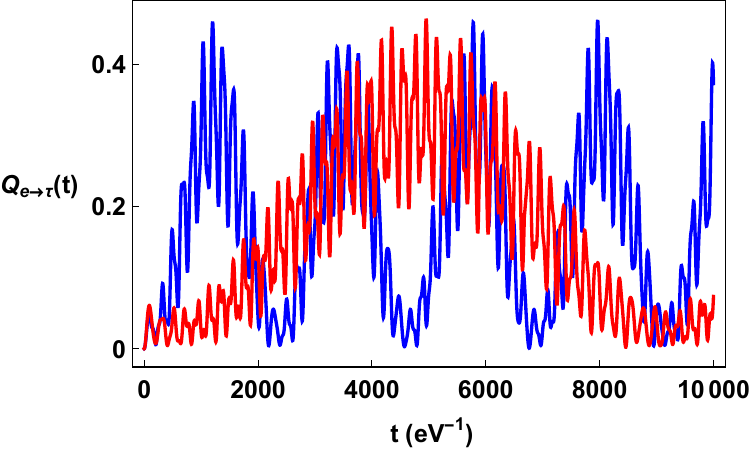}\hfill
\includegraphics[width=0.47\textwidth]{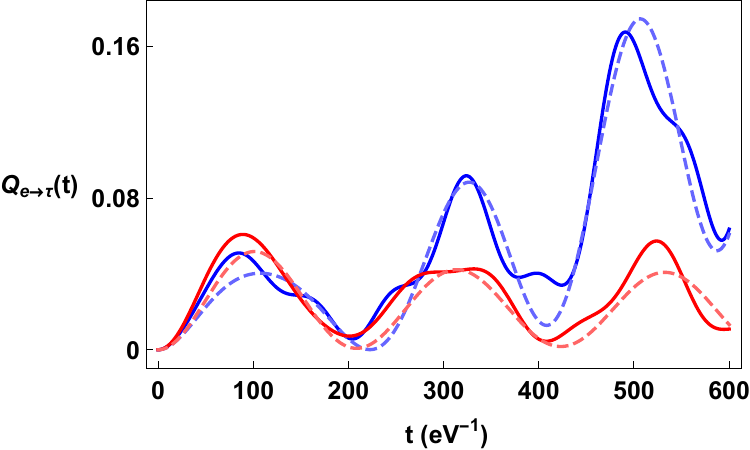}
\caption{Appearance channels in constant torsion. The upper row shows
$\nu_e\to\nu_\mu$ and the lower row $\nu_e\to\nu_\tau$. The left panels show
the full evolution for spin up (blue) and spin down (red); the right panels
show the short-time behavior and the quantum-mechanical comparison (dashed).}
\label{fig:constant-appearance}
\end{figure}

Figure~\ref{fig:constant-survival} compares the two spin sectors for electron
survival. The difference between the blue and red curves is substantial over
the full evolution and remains visible in the short-time comparison with the
quantum-mechanical result. The appearance channels in
Fig.~\ref{fig:constant-appearance} show the same qualitative behavior.

Three parametric regimes should be distinguished. For
$|T^3|\ll m_i,k$, the standard torsionless result is recovered. For a torsion
scale comparable to the masses and momenta, the splitting of both energies and
Bogoliubov amplitudes is largest. Finally, if $|T^3|\gg m_i,k$, the common
torsional contribution dominates the spectrum. The relative differences
between mass eigenstates then become small compared with the total energy,
reducing flavor conversion; at the same time, the two spin sectors become
approximately degenerate in their oscillation pattern. The large-torsion limit
therefore does not produce an arbitrarily large observable effect. It erases
the very mass splittings that drive oscillations.

\subsection{Time-dependent torsion}

\begin{figure}[htbp]
\centering
\includegraphics[width=0.47\textwidth]{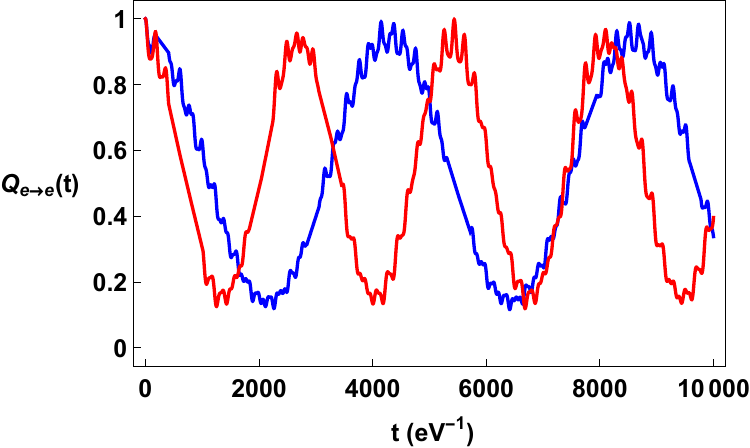}\hfill
\includegraphics[width=0.47\textwidth]{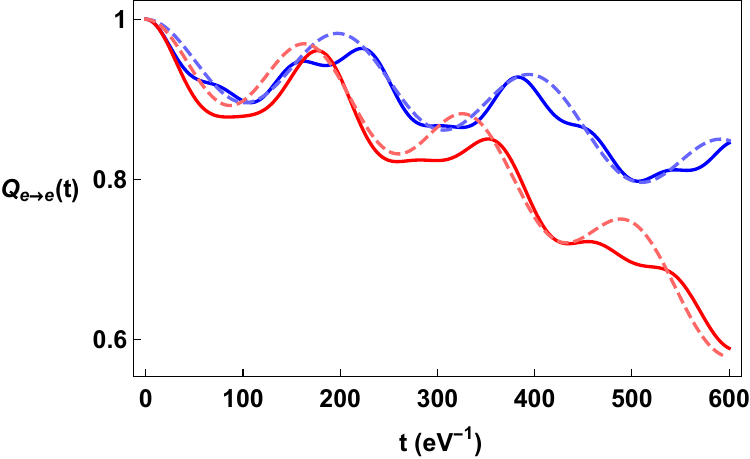}\\[0.8em]
\includegraphics[width=0.47\textwidth]{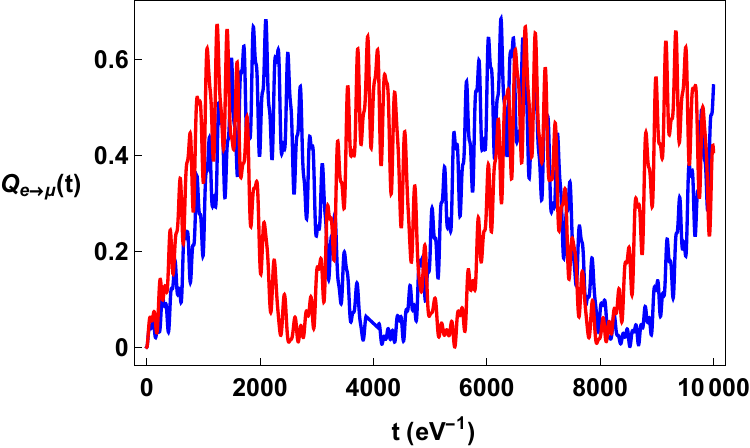}\hfill
\includegraphics[width=0.47\textwidth]{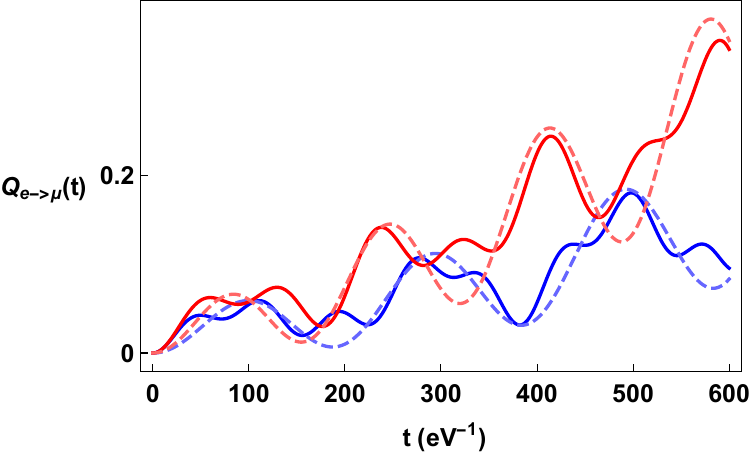}\\[0.8em]
\includegraphics[width=0.47\textwidth]{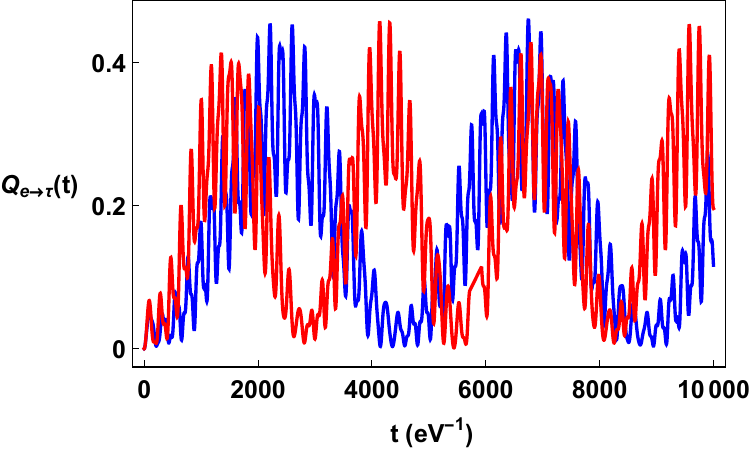}\hfill
\includegraphics[width=0.47\textwidth]{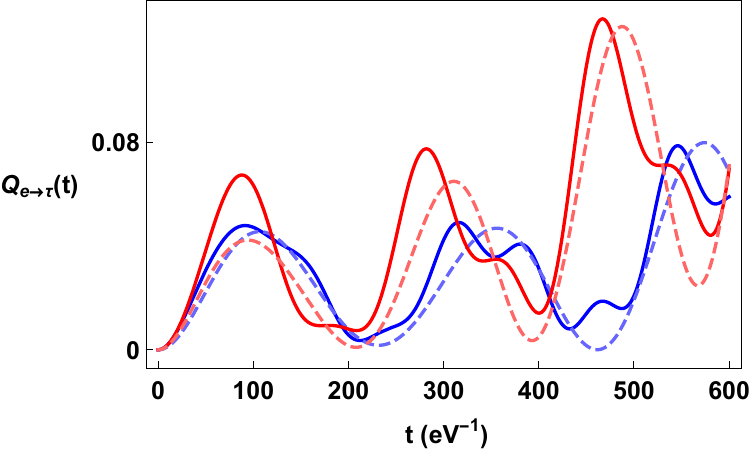}
\caption{Oscillation functions in a linearly time-dependent torsion
background for $\eta\breve T^0=5\times10^{-3}\,\mathrm{eV}$. From top to
bottom: $\nu_e\to\nu_e$, $\nu_e\to\nu_\mu$, and $\nu_e\to\nu_\tau$. Left
panels show the full evolution; right panels show the short-time comparison
with the corresponding quantum-mechanical curves.}
\label{fig:time-dependent}
\end{figure}

For the linearly varying spatial background, we retain the same neutrino
parameters and take $\eta\breve T^0=5\times10^{-3}\,\mathrm{eV}$. The
probabilities retain the structure of Eq.~\eqref{eq:survival}, with
$(\Gamma,\Sigma)=(\Pi,\Upsilon)$. Figure~\ref{fig:time-dependent} shows that the
spin dependence persists and is modulated by the nonlinear phase generated by
the spatial torsion profile.

The time-dependent case is more model dependent than the constant one. The
assumption that $\breve T^0$ is constant is what makes the evolution operator
analytically tractable, and the neglect of spin-flip transitions removes a
potentially important channel. The curves should therefore be interpreted as a
controlled demonstration of the phase and Bogoliubov effects, not as a
complete prediction for an arbitrary dynamical torsion source.

\subsection{$CP$ asymmetry and flavor-vacuum condensate}

\begin{figure}[htbp]
\centering
\includegraphics[width=0.47\textwidth]{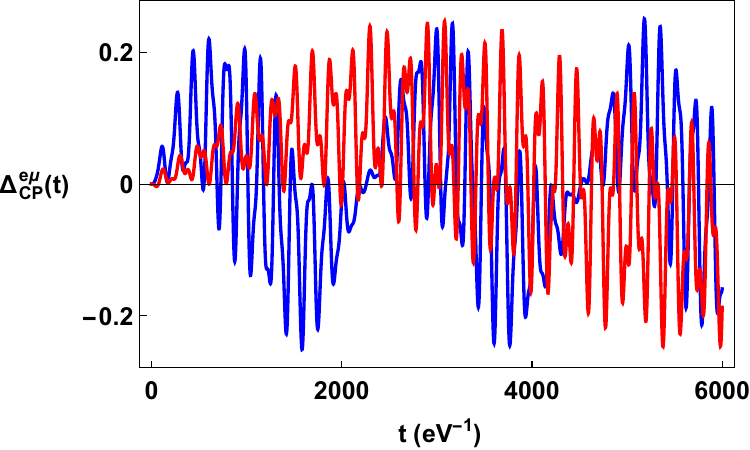}\hfill
\includegraphics[width=0.47\textwidth]{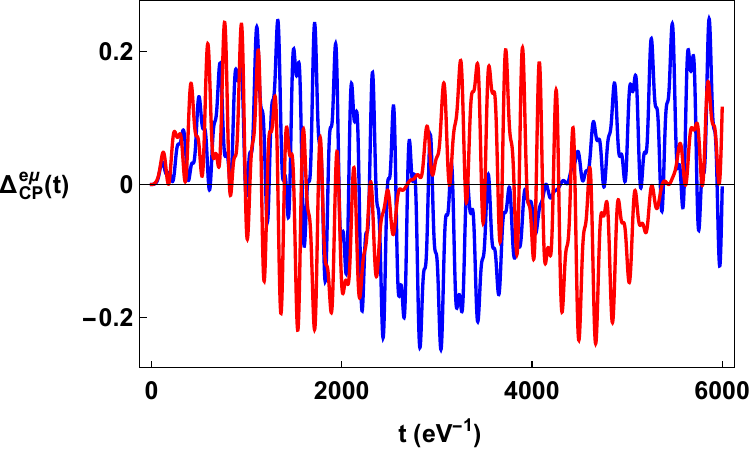}
\caption{Spin-dependent $CP$ asymmetry in the $\nu_e\to\nu_\mu$ channel.
The left panel refers to the constant-torsion benchmark, whereas the right
panel refers to the linearly time-dependent background. In both panels the
blue and red curves correspond to the two spin orientations.}
\label{fig:cp}
\end{figure}

For fixed spin $r$, the field-theoretic $CP$ asymmetry is defined by
\begin{equation}
\Delta_{r;CP}^{\rho\sigma}(t)
=\mathcal Q_{\nu_\rho\to\nu_\sigma}^{r,\bm k}(t)
+\mathcal Q_{\bar\nu_\rho\to\bar\nu_\sigma}^{r,\bm k}(t),
\label{eq:cpdef}
\end{equation}
where the plus sign reflects the negative flavor charge carried by the
antineutrino state. For $\nu_e\to\nu_\mu$ one obtains
\begin{align}
\Delta_{r;CP}^{e\mu}(t)
=4J_{CP}\Bigl\{&
|\Gamma_{12}^{rr}|^2\sin(2\Delta_{12}^rt)
-|\Sigma_{12}^{rr}|^2\sin(2\Omega_{12}^rt)
\nonumber\\
&+\left(|\Gamma_{12}^{rr}|^2-|\Sigma_{13}^{rr}|^2\right)
\sin(2\Delta_{23}^rt)
+\left(|\Sigma_{12}^{rr}|^2-|\Sigma_{13}^{rr}|^2\right)
\sin(2\Omega_{23}^rt)
\nonumber\\
&-|\Gamma_{13}^{rr}|^2\sin(2\Delta_{13}^rt)
+|\Sigma_{13}^{rr}|^2\sin(2\Omega_{13}^rt)
\Bigr\},
\label{eq:cpasymmetry}
\end{align}
where momentum labels have been suppressed and
\begin{equation}
J_{CP}=\frac{1}{8}\sin\delta\,
\sin(2\theta_{12})\sin(2\theta_{13})\cos\theta_{13}
\sin(2\theta_{23}).
\end{equation}
Because every frequency and coefficient in Eq.~\eqref{eq:cpasymmetry} depends
on $r$, torsion makes the $CP$ asymmetry spin dependent, as illustrated in
Fig.~\ref{fig:cp}.

\begin{figure}[htbp]
\centering
\includegraphics[width=0.47\textwidth]{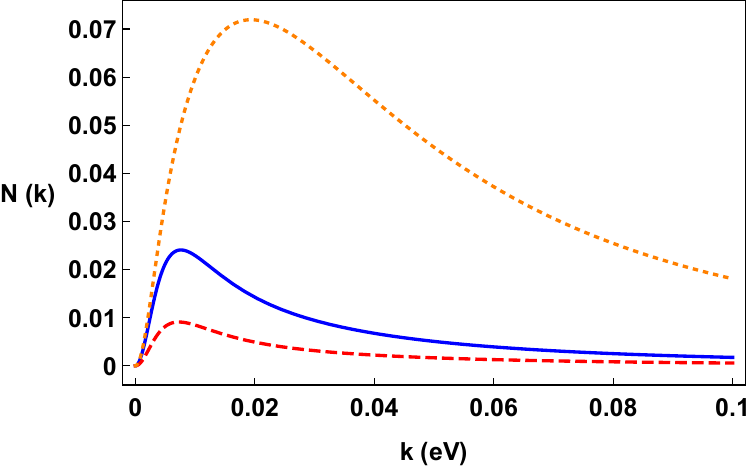}\hfill
\includegraphics[width=0.47\textwidth]{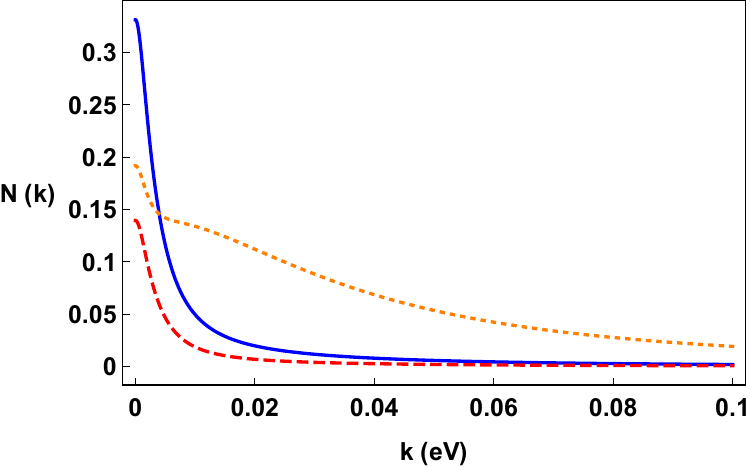}
\caption{Flavor-vacuum condensation densities as functions of momentum for
the constant-torsion benchmark. The left and right panels correspond to the
two spin orientations. In each panel the three curves represent
$\mathcal N_{1;\bm k}^{r}$ (blue solid),
$\mathcal N_{2;\bm k}^{r}$ (red dashed), and
$\mathcal N_{3;\bm k}^{r}$ (orange dotted).}
\label{fig:condensate}
\end{figure}

The same Bogoliubov transformation implies a nontrivial flavor-vacuum
condensate. For example,
\begin{equation}
\mathcal N_{1;\bm k}^{r}
={}_f\!\bra{0(t)}N_{\alpha_1,\bm k}^{r}\ket{0(t)}_f
=s_{12}^{2}c_{13}^{2}|\Sigma_{12;\bm k}^{rr}|^{2}
+s_{13}^{2}|\Sigma_{13;\bm k}^{rr}|^{2},
\label{eq:condensate}
\end{equation}
with analogous expressions for $i=2,3$. Since $\Sigma^{++}\neq\Sigma^{--}$,
the torsion background breaks the spin degeneracy of the condensation
densities. Their momentum dependence follows directly from the spin-dependent
coefficients in Eq.~\eqref{eq:condensate}, as shown in
Fig.~\ref{fig:condensate}.

The effect is strongest for nonrelativistic neutrinos, precisely where the
Bogoliubov coefficients differ most from their Pontecorvo limit. Experiments
sensitive to the cosmic neutrino background, such as PTOLEMY
\cite{Betti2019}, are therefore conceptually relevant. A realistic forecast,
however, requires substantially more than the benchmark curves shown here: one
must specify a dynamical source of torsion, its spatial profile, the spin and
momentum distribution of the neutrino flux, and the detector response. Without
that matching, the present results establish a distinctive mechanism but not a
quantitative experimental reach.

\section{Conclusions}

We have summarized the quantum-field-theoretic description of neutrino mixing
in constant and linearly time-dependent torsion backgrounds. In the
Einstein--Cartan setting, minimally coupled Dirac fields interact with the
axial component of torsion. A constant spatial background produces
spin-dependent effective masses and energies. When flavor mixing is treated at
the operator level, the same background also changes the spinor overlaps and
hence the Bogoliubov coefficients entering the flavor ladder operators.
Torsion therefore modifies both the frequencies and the amplitudes of the exact
oscillation formulae.

The resulting spin dependence is most pronounced when the torsion scale and
the neutrino masses are comparable and the momentum is low. In the opposite
limit of a dominant torsion term, the relative mass splittings become
subleading and flavor conversion is suppressed rather than enhanced. The
mechanism also produces spin-dependent $CP$ asymmetries and unequal condensate
densities for the two spin sectors of the flavor vacuum. These effects vanish
in the torsionless ultrarelativistic limit, where the Pontecorvo formulae are
recovered.

The analysis is intentionally kinematical. Torsion is prescribed as an
external background, curvature is neglected, and spin-flip transitions are
omitted in the time-dependent example. The next serious step is therefore not
to add more benchmark plots, but to embed the calculation in a specified
astrophysical or cosmological torsion model and propagate its predicted
profile through an experimentally realistic neutrino distribution. Only that
matching can determine whether the spin-dependent QFT signatures derived here
are observable.

\begin{acknowledgments}
Partial financial support from MUR and INFN is acknowledged. A.C. also
acknowledges the COST Action CA1511, Cosmology and Astrophysics Network for
Theoretical Advances and Training Actions (CANTATA).
\end{acknowledgments}

\bibliography{bibliografia}

@article{Capozziello2011,
    author = "Capozziello, Salvatore and De Laurentis, Mariafelicia",
    title = "{Extended Theories of Gravity}",
    eprint = "1108.6266",
    archivePrefix = "arXiv",
    primaryClass = "gr-qc",
    doi = "10.1016/j.physrep.2011.09.003",
    journal = "Phys. Rept.",
    volume = "509",
    pages = "167--321",
    year = "2011"
}

@article{BransDicke,
    author = "Brans, C. and Dicke, R. H.",
    editor = "Hsu, Jong-Ping and Fine, D.",
    title = "{Mach's principle and a relativistic theory of gravitation}",
    doi = "10.1103/PhysRev.124.925",
    journal = "Phys. Rev.",
    volume = "124",
    pages = "925--935",
    year = "1961"
}

@article{Quiros2019,
    author = "Quiros, Israel",
    title = "{Selected topics in scalar{\textendash}tensor theories and beyond}",
    eprint = "1901.08690",
    archivePrefix = "arXiv",
    primaryClass = "gr-qc",
    doi = "10.1142/S021827181930012X",
    journal = "Int. J. Mod. Phys. D",
    volume = "28",
    number = "07",
    pages = "1930012",
    year = "2019"
}

@article{Kobayashi2019,
    author = "Kobayashi, Tsutomu",
    title = "{Horndeski theory and beyond: a review}",
    eprint = "1901.07183",
    archivePrefix = "arXiv",
    primaryClass = "gr-qc",
    reportNumber = "RUP-19-3",
    doi = "10.1088/1361-6633/ab2429",
    journal = "Rept. Prog. Phys.",
    volume = "82",
    number = "8",
    pages = "086901",
    year = "2019"
}

@article{Hehl1976,
    author = "Hehl, F. W. and Von Der Heyde, P. and Kerlick, G. D. and Nester, J. M.",
    title = "{General Relativity with Spin and Torsion: Foundations and Prospects}",
    doi = "10.1103/RevModPhys.48.393",
    journal = "Rev. Mod. Phys.",
    volume = "48",
    pages = "393--416",
    year = "1976"
}

@article{Shapiro2002,
    author = "Shapiro, I. L.",
    title = "{Physical aspects of the space-time torsion}",
    eprint = "hep-th/0103093",
    archivePrefix = "arXiv",
    reportNumber = "DF-UFJF-01-04",
    doi = "10.1016/S0370-1573(01)00030-8",
    journal = "Phys. Rept.",
    volume = "357",
    pages = "113",
    year = "2002"
}

@article{Cabral2021,
    author = "Cabral, Francisco and Lobo, Francisco S. N. and Rubiera-Garcia, Diego",
    title = "{Imprints from a Riemann{\textendash}Cartan space-time on the energy levels of Dirac spinors}",
    eprint = "2102.02048",
    archivePrefix = "arXiv",
    primaryClass = "gr-qc",
    doi = "10.1088/1361-6382/ac1cca",
    journal = "Class. Quant. Grav.",
    volume = "38",
    number = "19",
    pages = "195008",
    year = "2021"
}

@article{Cirilo2019,
    author = "Julio Cirilo-Lombardo, Diego",
    title = "{Fermion helicity flip and fermion oscillation induced by dynamical torsion field}",
    doi = "10.1209/0295-5075/127/10002",
    journal = "EPL",
    volume = "127",
    number = "1",
    pages = "10002",
    year = "2019"
}

@article{Bilenky1978,
    author = "Bilenky, Samoil M. and Pontecorvo, B.",
    title = "{Lepton Mixing and Neutrino Oscillations}",
    doi = "10.1016/0370-1573(78)90095-9",
    journal = "Phys. Rept.",
    volume = "41",
    pages = "225--261",
    year = "1978"
}

@article{Bilenky1987,
    author = "Bilenky, Samoil M. and Petcov, S. T.",
    title = "{Massive Neutrinos and Neutrino Oscillations}",
    doi = "10.1103/RevModPhys.59.671",
    journal = "Rev. Mod. Phys.",
    volume = "59",
    pages = "671",
    year = "1987",
    note = "[Erratum: Rev.Mod.Phys. 61, 169 (1989), Erratum: Rev.Mod.Phys. 60, 575--575 (1988)]"
}

@article{Fukuda1998,
    author = "Fukuda, Y. and others",
    collaboration = "Super-Kamiokande",
    title = "{Evidence for oscillation of atmospheric neutrinos}",
    eprint = "hep-ex/9807003",
    archivePrefix = "arXiv",
    reportNumber = "BU-98-17, ICRR-REPORT-422-98-18, UCI-98-8, KEK-PREPRINT-98-95, LSU-HEPA-5-98, UMD-98-003, SBHEP-98-5, TKU-PAP-98-06, TIT-HPE-98-09",
    doi = "10.1103/PhysRevLett.81.1562",
    journal = "Phys. Rev. Lett.",
    volume = "81",
    pages = "1562--1567",
    year = "1998"
}

@article{Blasone1995,
    author = "Blasone, M. and Vitiello, Giuseppe",
    title = "{Quantum field theory of fermion mixing}",
    eprint = "hep-ph/9501263",
    archivePrefix = "arXiv",
    reportNumber = "SADF1-1995",
    doi = "10.1006/aphy.1995.1115",
    journal = "Annals Phys.",
    volume = "244",
    pages = "283--311",
    year = "1995",
    note = "[Erratum: Annals Phys. 249, 363--364 (1996)]"
}

@article{Blasone2002,
    author = "Blasone, Massimo and Capolupo, Antonio and Vitiello, Giuseppe",
    title = "{Quantum field theory of three flavor neutrino mixing and oscillations with CP violation}",
    eprint = "hep-th/0204184",
    archivePrefix = "arXiv",
    doi = "10.1103/PhysRevD.66.025033",
    journal = "Phys. Rev. D",
    volume = "66",
    pages = "025033",
    year = "2002"
}

@article{Capolupo2016,
    author = "Capolupo, Antonio",
    title = "{Dark matter and dark energy induced by condensates}",
    eprint = "1608.05407",
    archivePrefix = "arXiv",
    primaryClass = "hep-th",
    doi = "10.1155/2016/8089142",
    journal = "Adv. High Energy Phys.",
    volume = "2016",
    pages = "8089142",
    year = "2016"
}

@article{Adak2001,
    author = "Adak, Muzaffer and Dereli, T. and Ryder, L. H.",
    title = "{Neutrino oscillations induced by space-time torsion}",
    eprint = "gr-qc/0103046",
    archivePrefix = "arXiv",
    doi = "10.1088/0264-9381/18/8/307",
    journal = "Class. Quant. Grav.",
    volume = "18",
    pages = "1503--1512",
    year = "2001"
}

@article{Blasone2001,
    author = "Blasone, Massimo and Capolupo, Antonio and Romei, Oreste and Vitiello, Giuseppe",
    title = "{Quantum field theory of boson mixing}",
    eprint = "hep-ph/0102048",
    archivePrefix = "arXiv",
    doi = "10.1103/PhysRevD.63.125015",
    journal = "Phys. Rev. D",
    volume = "63",
    pages = "125015",
    year = "2001"
}

@article{Fabbri2016,
    author = "Fabbri, Luca and Vignolo, Stefano",
    title = "{A torsional completion of gravity for Dirac matter fields and its applications to neutrino oscillations}",
    eprint = "1504.03545",
    archivePrefix = "arXiv",
    primaryClass = "gr-qc",
    doi = "10.1142/S0217732316500140",
    journal = "Mod. Phys. Lett. A",
    volume = "31",
    number = "03",
    pages = "1650014",
    year = "2016"
}

@article{Betti2019,
    author = "Betti, M. G. and others",
    collaboration = "PTOLEMY",
    title = "{Neutrino physics with the PTOLEMY project: active neutrino properties and the light sterile case}",
    eprint = "1902.05508",
    archivePrefix = "arXiv",
    primaryClass = "astro-ph.CO",
    doi = "10.1088/1475-7516/2019/07/047",
    journal = "JCAP",
    volume = "07",
    pages = "047",
    year = "2019"
}

@article{CapozzielloVignolo2010,
  author  = {Capozziello, Salvatore and Vignolo, Stefano},
  title   = {Metric-affine {$f(R)$}-gravity with torsion: An overview},
  journal = {Annalen der Physik},
  volume  = {19},
  pages   = {238--248},
  year    = {2010},
  doi     = {10.1002/andp.201010420},
  eprint  = {0910.5230},
  archivePrefix = {arXiv},
  primaryClass  = {gr-qc}
}

@article{Cai2016,
  author  = {Cai, Yi-Fu and Capozziello, Salvatore and
             De Laurentis, Mariafelicia and Saridakis, Emmanuel N.},
  title   = {{$f(T)$} teleparallel gravity and cosmology},
  journal = {Reports on Progress in Physics},
  volume  = {79},
  number  = {10},
  pages   = {106901},
  year    = {2016},
  doi     = {10.1088/0034-4885/79/10/106901},
  eprint  = {1511.07586},
  archivePrefix = {arXiv},
  primaryClass  = {gr-qc}
}

@article{CapozzielloDeFalcoFerrara2022,
  author  = {Capozziello, Salvatore and De Falco, Vittorio and
             Ferrara, Carmen},
  title   = {Comparing equivalent gravities: Common features and differences},
  journal = {European Physical Journal C},
  volume  = {82},
  pages   = {865},
  year    = {2022},
  doi     = {10.1140/epjc/s10052-022-10823-x}
}

@article{CapolupoLambiaseQuaranta2020,
  author  = {Capolupo, Antonio and Lambiase, Gaetano and Quaranta, Aniello},
  title   = {Neutrinos in curved spacetime: Particle mixing and flavor oscillations},
  journal = {Physical Review D},
  volume  = {101},
  pages   = {095022},
  year    = {2020},
  doi     = {10.1103/PhysRevD.101.095022},
  eprint  = {2003.00516},
  archivePrefix = {arXiv},
  primaryClass  = {hep-th}
}

@article{CapolupoQuarantaSerao2023,
  author  = {Capolupo, Antonio and Quaranta, Aniello and Serao, Raoul},
  title   = {Field Mixing in Curved Spacetime and Dark Matter},
  journal = {Symmetry},
  volume  = {15},
  number  = {4},
  pages   = {807},
  year    = {2023},
  doi     = {10.3390/sym15040807}
}

@article{CapolupoCarloniQuaranta2022,
  author  = {Capolupo, Antonio and Carloni, Sante and Quaranta, Aniello},
  title   = {Quantum flavor vacuum in the expanding universe:
             A possible candidate for cosmological dark matter?},
  journal = {Physical Review D},
  volume  = {105},
  pages   = {105013},
  year    = {2022},
  doi     = {10.1103/PhysRevD.105.105013}
}

@article{CapolupoQuaranta2023,
  author  = {Capolupo, Antonio and Quaranta, Aniello},
  title   = {Neutrino capture on tritium as a probe of flavor vacuum
             condensate and dark matter},
  journal = {Physics Letters B},
  volume  = {839},
  pages   = {137776},
  year    = {2023},
  doi     = {10.1016/j.physletb.2023.137776},
  eprint  = {2205.09640},
  archivePrefix = {arXiv},
  primaryClass  = {hep-ph}
}

@article{CapolupoCapozzielloPisacaneQuaranta2025,
  author  = {Capolupo, Antonio and Capozziello, Salvatore and
             Pisacane, Gabriele and Quaranta, Aniello},
  title   = {Missing matter in galaxies as a neutrino mixing effect},
  journal = {Physics of the Dark Universe},
  volume  = {48},
  pages   = {101894},
  year    = {2025},
  doi     = {10.1016/j.dark.2025.101894},
  eprint  = {2411.17319},
  archivePrefix = {arXiv},
  primaryClass  = {hep-ph}
}

@article{CapolupoMondaPisacaneQuarantaSerao2025,
  author  = {Capolupo, Antonio and Monda, Simone and Pisacane, Gabriele
             and Quaranta, Aniello and Serao, Raoul},
  title   = {Dark Universe from QFT Mechanisms and Possible Experimental Probes},
  journal = {Universe},
  volume  = {11},
  number  = {5},
  pages   = {142},
  year    = {2025},
  doi     = {10.3390/universe11050142}
}

@article{CapolupoDeMariaMondaQuarantaSerao2024,
  author  = {Capolupo, Antonio and De Maria, Giuseppe and Monda, Simone
             and Quaranta, Aniello and Serao, Raoul},
  title   = {Quantum Field Theory of Neutrino Mixing in Spacetimes with Torsion},
  journal = {Universe},
  volume  = {10},
  number  = {4},
  pages   = {170},
  year    = {2024},
  doi     = {10.3390/universe10040170}
}

@article{CapolupoQuarantaSerao2025X17,
    author = "Capolupo, Antonio and Quaranta, Aniello and Serao, Raoul",
    title = "{The impact of the \(X_{17}\) boson on particle physics anomalies: Muon anomalous magnetic moment, Lamb shift and \(W\) mass}",
    eprint = "2410.01430",
    archivePrefix = "arXiv",
    primaryClass = "hep-ph",
    doi = "10.1016/j.dark.2024.101748",
    journal = "Phys. Dark Univ.",
    volume = "47",
    pages = "101748",
    year = "2025"
}

@article{SeraoTorreCapolupo2025QuantumInformation,
    author = "Serao, Raoul and Torre, Gianpaolo and Capolupo, Antonio",
    title = "{Quantum information meets high-energy physics: probing neutrinos and beyond}",
    eprint = "2510.22625",
    archivePrefix = "arXiv",
    primaryClass = "hep-ph",
    doi = "10.1088/1742-5468/ae1df3",
    journal = "J. Stat. Mech.",
    volume = "2025",
    number = "12",
    pages = "124001",
    year = "2025"
}

@article{CapolupoPisacaneQuarantaSerao2026Interferometry,
    author = "Capolupo, Antonio and Pisacane, Gabriele and Quaranta, Aniello and Serao, Raoul",
    title = "{Single arm interferometry to probe the scalar field dark matter}",
    eprint = "2505.13574",
    archivePrefix = "arXiv",
    primaryClass = "hep-ph",
    doi = "10.1103/ypm5-jh2j",
    journal = "Phys. Rev. D",
    volume = "113",
    number = "5",
    pages = "055027",
    year = "2026"
}

@article{CapolupoMondaPisacaneQuarantaSerao2026Torsion,
  author        = {Capolupo, Antonio and Monda, Simone and Pisacane, Gabriele and Quaranta, Aniello and Serao, Raoul},
  title         = {Spacetime torsion signatures in neutrino oscillation physics},
  journal       = {International Journal of Modern Physics A},
  year          = {2026},
  doi           = {10.1142/S0217751X26460139},
  eprint        = {2606.00617},
  archivePrefix = {arXiv},
  primaryClass  = {hep-th}
}

@article{CapozzielloHarkoKoivistoLoboOlmo2015,
  author        = {Capozziello, Salvatore and Harko, Tiberiu and Koivisto, Tomi S. and Lobo, Francisco S. N. and Olmo, Gonzalo J.},
  title         = {Hybrid Metric-Palatini Gravity},
  journal       = {Universe},
  volume        = {1},
  number        = {2},
  pages         = {199--238},
  year          = {2015},
  doi           = {10.3390/universe1020199},
  eprint        = {1508.04641},
  archivePrefix = {arXiv},
  primaryClass  = {gr-qc}
}

@article{CapozzielloCesareFerrara2025,
  author        = {Capozziello, Salvatore and Cesare, Sara and Ferrara, Carmen},
  title         = {Extended geometric trinity of gravity},
  journal       = {European Physical Journal C},
  volume        = {85},
  pages         = {932},
  year          = {2025},
  doi           = {10.1140/epjc/s10052-025-14440-2},
  eprint        = {2503.08167},
  archivePrefix = {arXiv},
  primaryClass  = {gr-qc}
}

@article{ManciniTinoCapozziello2025,
  author        = {Mancini, Christian and Tino, Guglielmo Maria and Capozziello, Salvatore},
  title         = {Equivalent Gravities and Equivalence Principle: Foundations and Experimental Implications},
  journal       = {Foundations of Physics},
  volume        = {55},
  number        = {5},
  pages         = {69},
  year          = {2025},
  doi           = {10.1007/s10701-025-00882-x},
  eprint        = {2501.06487},
  archivePrefix = {arXiv},
  primaryClass  = {gr-qc}
}
\end{document}